\pdfoutput=1 
\documentclass[letterpaper, 10 pt, conference,onecolumn]{ieeeconf} 

\IEEEoverridecommandlockouts             
\newcommand {\EDS}[1]{\mbox{ }\medskip \\ \textcolor{red}{\bf ***** {#1} *****} \mbox{ }\medskip \\}

\newcommand {\TODO}[1]{\mbox{ }\medskip \\ \textcolor{red}{\bf ***** TO DO: {#1} *****} \mbox{ }\medskip \\}

\renewcommand{\TODO}[1]{}

\newcommand{\tot}{{\textsc{t}}}
\newcommand{\subfxp}{{\scriptscriptstyle f(x,p)}}
\usepackage{amsmath,amssymb,graphicx,chemarr,xcolor}
\newcommand{\FIGURES}{FIGURES}
\newcommand{\pic}[2]{\includegraphics[width=#1\columnwidth]{\FIGURES/#2}}
\newcommand{\picc}[2]{\begin{center}\pic{#1}{#2}\end{center}}
\newtheorem{theorem}{Theorem}
\newtheorem{remark}{Remark}
\newtheorem{corollary}{Corollary}[theorem]
\newtheorem{claim}{Claim}
\usepackage{comment}

\newcommand{\beqn}{\begin{eqnarray*}}
\newcommand{\eeqn}{\end{eqnarray*}}
\newcommand{\halmos}{\rule{1ex}{1.4ex}}
\newenvironment{myproof}{\noindent {\em Proof}.\ }{\hspace*{\fill}$\halmos$\medskip}

\newcommand{\epr}{\end{myproof}}
\newcommand{\bpr}{\begin{myproof}}

\usepackage[utf8]{inputenc}
\usepackage{listings,xcolor}

\title{\LARGE \bf
Competition for binding targets results in paradoxical effects for simultaneous activator and repressor action - Extended Version
}

\author{M. Ali Al-Radhawi$^{1,*}$,
  Krishna Manoj$^{2,*}$,
  Dhruv D. Jatkar$^{3}$,
  Alon Duvall$^{4}$,\\
Domitilla Del Vecchio$^{5}$,
Eduardo D. Sontag$^{6}$
\thanks{This work was partially supported by grants
AFOSR FA9550-22-1-0316 and NSF/DMS-2052455}%
\thanks{$^{*}$Co-first authors}%
\thanks{$^{1}$Northeastern University
{\tt\small malirdwi@gmail.com}}%
\thanks{$^{2}$MIT
{\tt\small kmanoj@mit.edu }}%
\thanks{$^{3}$Northeastern University
{\tt\small jatkar.d@northeastern.edu}}%
\thanks{$^{4}$Northeastern University
{\tt\small duvall.a@northeastern.edu}}%
\thanks{$^{5}$MIT
{\tt\small ddv@mit.edu}}%
\thanks{$^{6}$Northeastern University
{\tt\small e.sontag@northeastern.edu}}
}
\begin{document}

\maketitle
\thispagestyle{empty}
\pagestyle{empty}

\begin{abstract}

In the context of epigenetic transformations in cancer metastasis, a puzzling effect was recently discovered, in which the elimination (knock-out) of an activating regulatory element leads to increased (rather than decreased) activity of the element being regulated. It has been postulated that this paradoxical behavior can be explained by activating and repressing transcription factors competing for binding to other possible targets. It is very difficult to prove this hypothesis in mammalian cells, due to the large number of potential players and the complexity of endogenous intracellular regulatory networks. Instead, this paper analyzes this issue through an analogous synthetic biology construct which aims to reproduce the paradoxical behavior using standard bacterial gene expression networks. The paper first reviews the motivating cancer biology work, and then describes a proposed synthetic construct. A mathematical model is formulated, and basic properties of uniqueness of steady states and convergence to equilibria are established, as well as an identification of parameter regimes which should lead to observing such paradoxical phenomena (more activator leads to less activity at steady state). A proof is also given to show that this is a steady-state property, and for initial transients the phenomenon will not be observed.
This work adds to the general line of work of resource competition in synthetic circuits.
\end{abstract}

\section{INTRODUCTION AND BACKGROUND}

The field of synthetic biology has as its ultimate goal to program new or modify existing biological systems, for applications ranging from cell therapies and regenerative medicine to biosensing and biofuel production.  In general, a significant obstacle to the development of synthetic biological circuits is the influence of compositional context: the fundamental characteristics of a circuit alter in the presence of additional components due to competition for resources, off-target interactions, genetic context, growth rate feedback loops, and retroactivity effects.
See e.g.
\cite{%
ddv07,
2012_cardinale_arkin,%
SHAKIBA2021561,%
MAli_competition_phenotypes,%
DELVECCHIO2015111,%
STONE2024%
}
and
\cite{delvecchio_qian_murray_sontag_annual_reviews2018} for an overview.
Unless one designs mathematically-validated control circuits to compensate for uncertainty, designers will need to re-adjust each part whenever new elements are integrated into a system.


In this paper, we study a synthetic design that aims to validate the competition principles in a model from \cite{MAli_EMT_2022} that has been hypothesised to explain a paradoxical effect in cell differentiation experiments.
The transformation of genetically identical cells into distinct phenotypes, and the transitions between these phenotypes, are governed by complex processes involving epigenetic markers as well as more classical gene-regulatory networks (GRNs) involving transcription factors and non-coding regulatory RNAs. In living cells, such interactions are complicated by the potential competition of multiple TFs over one target, and also by the sequestration of a TF by multiple targets. However, it is not entirely clear whether such effects are able to drive cellular decision making. In a recent publication~\cite{MAli_EMT_2022}, we have hypothesised that the competition for genomic targets among epigenetic factors can provide an explanation for puzzling experimental data regarding the epithelial–mesenchymal transition (EMT) in cancer metastasis. 
Our mathematical analysis predicted that when the activity of a regulator is perturbed, it can lead to the widespread redistribution of epigenetic marks, thereby influencing the levels of competing regulators. 

Since it is hard to test this mechanistic hypothesis through epigenetic modifications in mammalian cells, we propose here a synthetic biology analog involving transcription factors in bacterial cells. In this paper, we review the mechanism and make mathematical predictions from a model, as well as a proposed implementation using CRISPRi/a technology. Experimental work is ongoing.
In the remainder of this section, we review our previous results \cite{MAli_EMT_2022}.

\subsection{Review of previous work}
\begin{figure}[ht]
\picc{0.45}{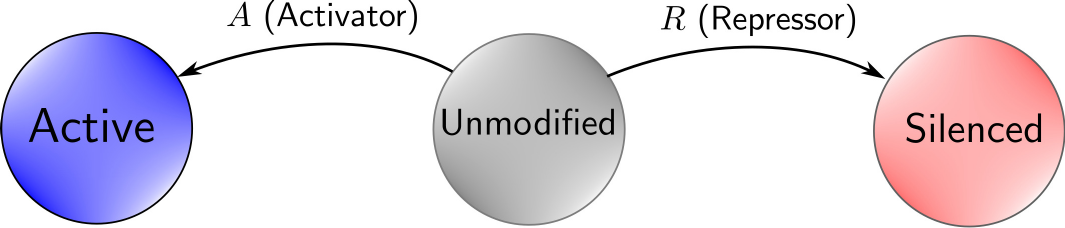}
\caption{A simplified model of a target that can occupy three different states.}
\label{fig:overlapping_promoter}
\end{figure}

\noindent{\bf Regulatory factors competing for the same target.} Depending on the nature of a regulator, its target can be a promoter, a histone tail, an mRNA, or others. However, mathematically, we can represent the state of a given target using the same simplified three-state model depicted in 
Figure~\ref{fig:overlapping_promoter}. 
If a target has not been subject to the activity of a regulator, it retains its nominal state which we call ``unmodified''. A regulator can change the nominal state into ``active'' or ``silenced'' depending on whether it is an activator or repressor, respectively.  If the a target is subject to the activity of regulators with opposing effects, then it can switch between the three different states depending on the binding affinities and the abundances of the regulators. 

\noindent{\bf Motivation: Experiments on a cancer cell line.} Epigenetic regulation has many examples in which opposing regulators compete for the same targets. A prominent example is the antagonism between the Polycomb complex group (PcG) and the Trithorax (TrX) group  \cite{schuettengruber2017genome}.  This system has been probed in a recent set of knockout experiments in a breast cancer cell line \cite{zhang2022genome} using CRISPR. 
The knockout of PRC2 (a PcG repressor) and KMT2D (a TrX activator) initiated two distinct trajectories of epithelial-to-mesenchymal transitions (EMTs). Using the language of dynamical systems, the system settled into two different steady states depending on the particular perturbation. 

\begin{figure}[t]
\picc{0.65}{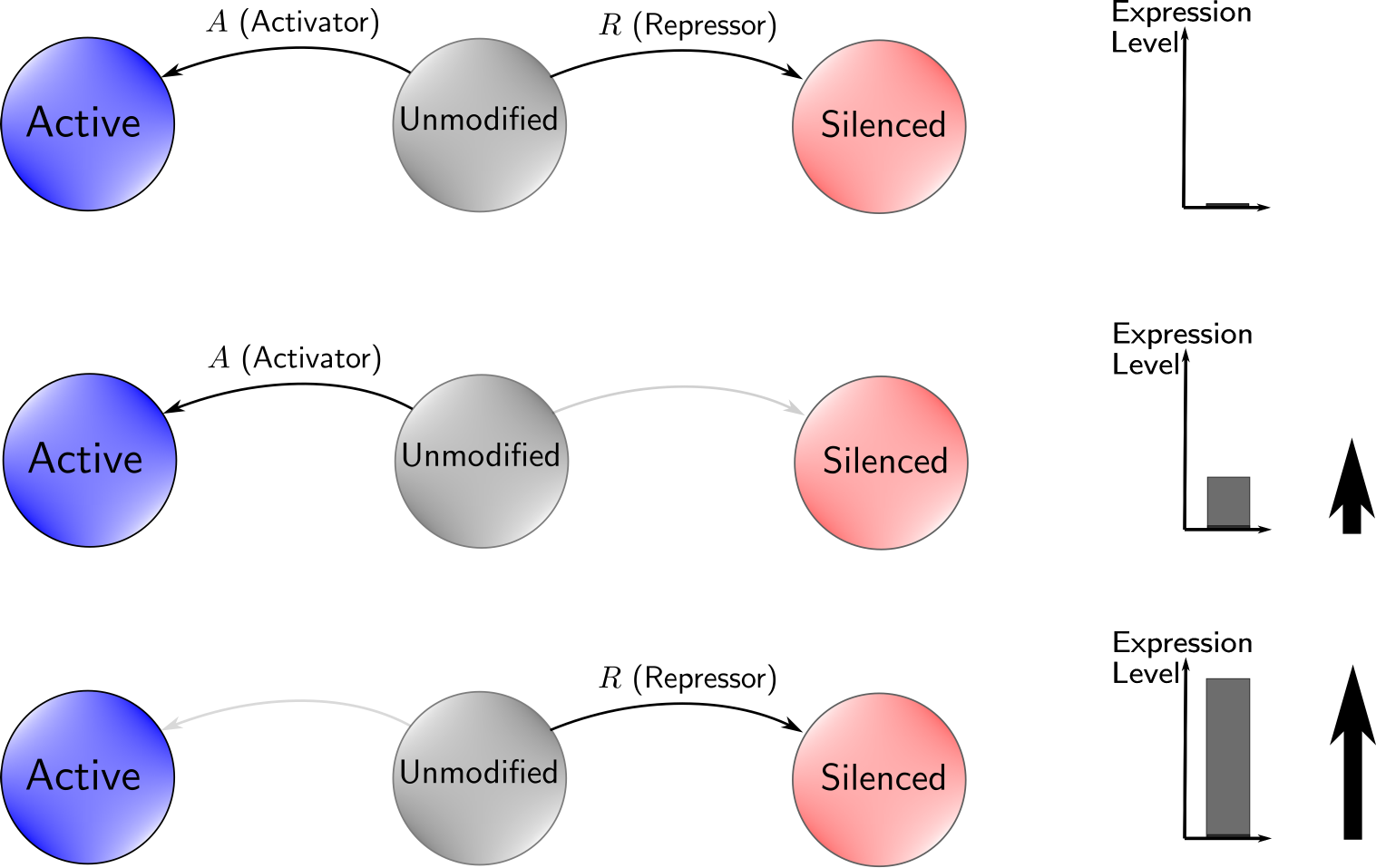}
\caption{Paradoxical results of the knockout experiments from \cite{zhang2022genome}. Knockout of a repressor is not able to restore the full activity of a target gene (e.g, ZEB1), while knocking out an activator results in maximal activation.}
\label{fig:yun_results}
\end{figure}

\noindent{\bf Motivation: Paradoxical gene activity pattern.}
In the aforementioned experiments, a paradoxical pattern of gene activation was observed that cannot be easily explained by known local gene regulatory interactions. Consider a gene of interest (e.g. ZEB1 which is a major driver of EMT). As shown in  Figure~\ref{fig:yun_results},
the gene is nominally repressed due to the dominant activity of the repressor (PRC2). However, when the repressor is knocked out, the gene achieves a mediocre amount of activation. In the second knockout experiment, an activating regulator (KMT2D) is knocked out. In that case, the gene achieves  maximal activation despite the fact that its main repressor is not knocked-out. The main question is: how would the direct knockout of a repressing regulator be \emph{less effective} in activating a gene compared to knocking out an activator? We summarize our answer next.
 
\noindent{\bf Our model: off-target competition causes sequestration.}
In our recent paper \cite{MAli_EMT_2022}, we reviewed the relevant literature on PcG/TrX regulators and distilled it into four postulates:
\begin{enumerate}
 \item Regulators compete for binding to the same targets. 
 \item There is a large number of targets per regulator, e.g.  hundreds or thousands. 
 \item  Each regulator has limited levels. 
 \item When a regulator molecule is active at a given target, it cannot influence the activity of other targets.
\end{enumerate}

\begin{figure}[t]
	\picc{0.75}{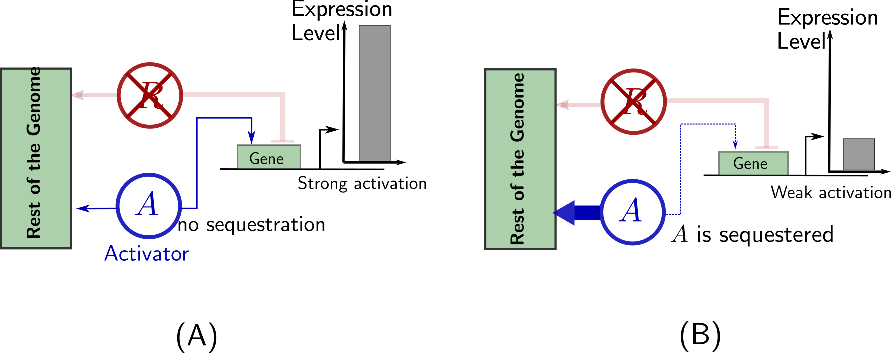}
	\caption{Repressor knockout case. (a) global context has no effect. (b) global context has considerable effect.}
	\label{fig:caseIIcontrolTF}
\end{figure}
\begin{figure}[t]
\picc{0.75}{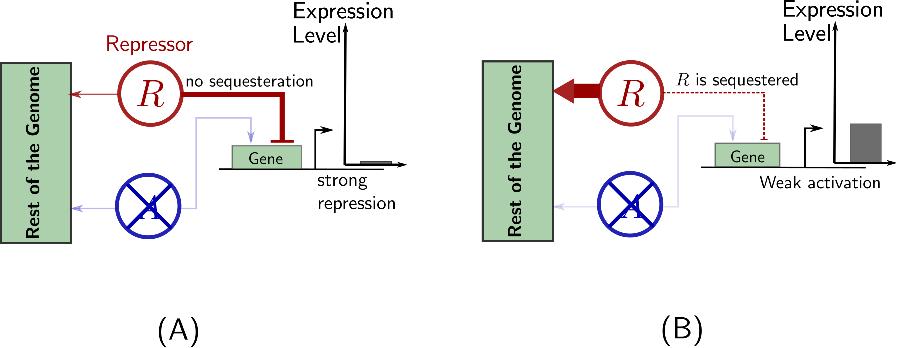}
\caption{Activator knockout case. (A) global context has no effect. (B) global context has considerable effect.}
\label{fig:caseIIKKOTF}
\end{figure}

\noindent{\bf Toy model with two factors.}
Consider a toy model of two regulators competing for a target as shown in  Figure~\ref{fig:caseIIcontrolTF}. 
In the absence of regulators, the target is assumed to be weakly active. When the regulators are present, assume that the repressor is dominant and it is able to robustly silence the target. Let us consider an experiment in which the repressor in knocked out. In a situation in which there is minimal off-target interference, we expect that the activator utilizes the absence of its competitor to strongly activate its target as shown in Figure~\ref{fig:caseIIcontrolTF}-a). However, assume now that the activator shares many other targets with the repressor across the genome, and once the repressor knockout, a ``void'' is created across the genome, and the activator has too many potential targets. Depending on the relative affinities, the activator can get sequestered into other targets across the genome leaving out the target gene without activation as shown in Figure~\ref{fig:caseIIcontrolTF}-b).  The opposite scenario can be similarly illustrated as in Figure~\ref{fig:caseIIKKOTF} where the repressor is still nominally dominant. When the activator is knocked out, the repressor can silence the target when there is minimal off-target interference as shown in Figure~\ref{fig:caseIIKKOTF}-a). 
However, the repressor can get sequestered to other off-targets when there is significant affinity to off-targets as shown in Figure~\ref{fig:caseIIKKOTF}-b), and hence an activator can indirectly activate a target gene by its absence.  

This simplified model can partially explain the paradoxical results shown in Figure~\ref{fig:yun_results}.~It shows how repressor knockout can fail to activate a target gene, and how an activator knockout can activate the target. However, it does not show how can an activator knockout be more effective at activation than a repressor knockout. This is since the mechanism of activation depends on the sequestration of the repressor, hence it cannot yield an activation that is stronger than a full repressor knockout. 
The paper~\cite{MAli_EMT_2022} shows how to obtain the full spectrum of behaviors using three factors: one repressor and two activators of different strengths.
On the other hand, one can obtain all three cases in Figure~\ref{fig:yun_results} provided that partial knock-outs (i.e., ``knock-downs'') of genes are possible.

As explained in the introduction, it is very difficult to test this mechanistic explanation of the paradoxical effects through epigenetic modifications in mammalian cells. Thus, in this work, we describe and analyze a synthetic biology knock-down model with two factors that uses transcription factors in bacterial cells. 


\section{Proposed synthetic construct}

\begin{figure}[ht]
\picc{0.45}{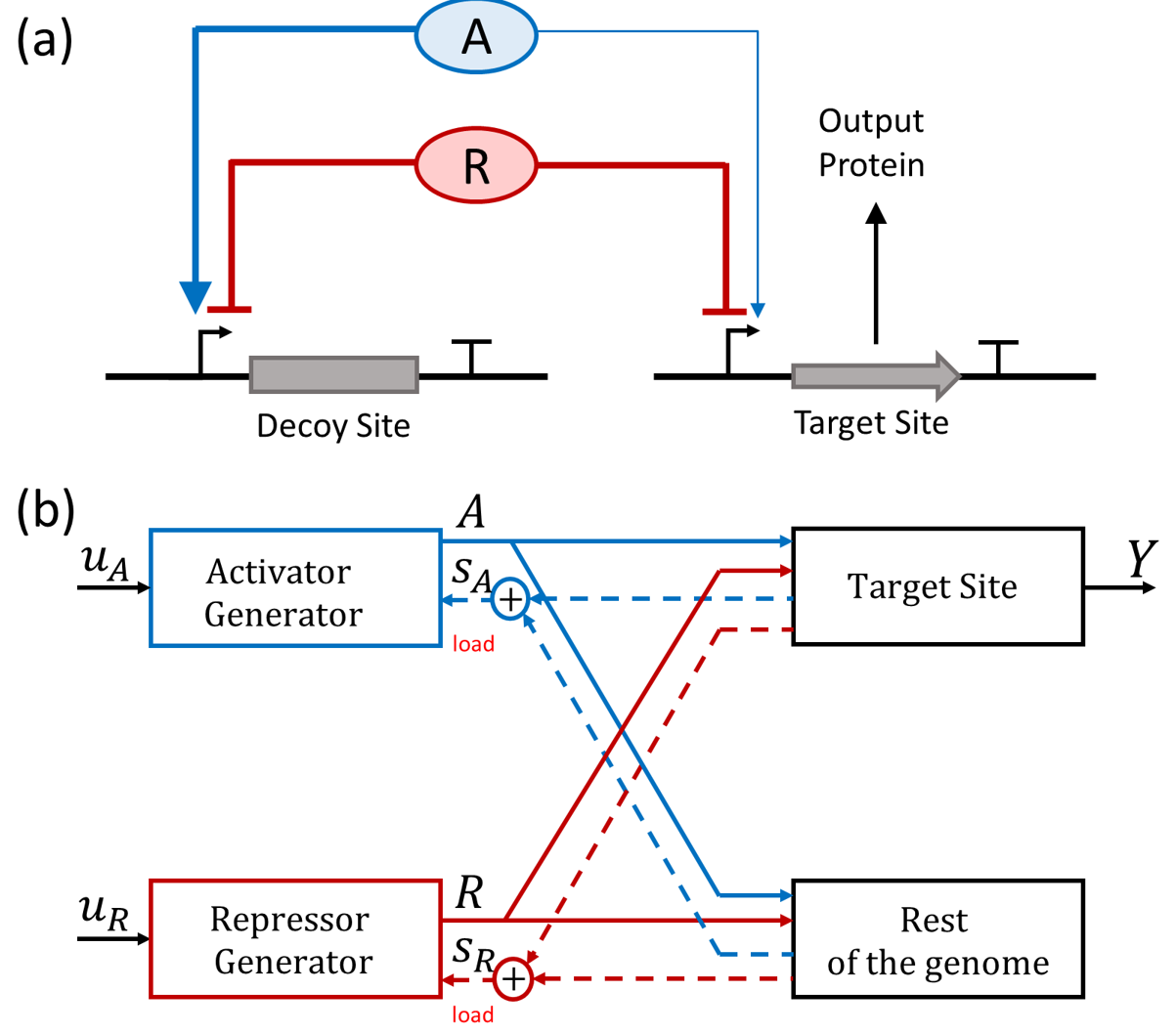}
\caption{Proposed project: a synthetic competition circuit. (a) Genetic circuit diagram and (b) block diagram representation of the proposed system. Solid lines denote the intended input and outputs and the dashed lines display the unintended loads due to competition in the circuit.}
\label{fig:gencir_bd}
\end{figure}

The proposed circuit (shown in Figure~\ref{fig:gencir_bd}) consists of a pool of shared resources (activators and repressors) regulating the production of the output protein by the target gene while competing with the rest of the genome, modeled as decoy sites.
The transcription factors, activator ($A$, produced at a constant rate $u_A$) and repressor ($R$, produced at a constant rate $u_R$) bind to the target sites ($T$) to transform into an active form ($T_A$) and silenced form ($T_R$). The active form of the target undergoes transcription and translation to produce the output protein ($Y$) at a rate $\kappa > \beta$, where $\beta$ is the basal transcription rate in the inactive form of the target. The decoy sites ($D$) sequester the resources by forming $D_A$ and $D_R$ complexes respectively. The chemical reactions involved are: 
%
%
%
%
%
\begin{align}
\emptyset & \xrightleftharpoons[\Bar{\delta}]{u_A} A  &\emptyset & \xrightleftharpoons[\Bar{\delta}]{u_R} R \nonumber\\
R+T & \xrightleftharpoons[t_r^-]{t_r^+} T_{R},  &A+T & \xrightleftharpoons[t_a^-]{t_a^+} T_{A} ~~\mbox{(Target)} \nonumber\\
R+D & \xrightleftharpoons[d_r^-]{d_r^+} D_{R} &A+D & \xrightleftharpoons[d_a^-]{d_a^+} D_{A} ~\mbox{(Decoy)} \nonumber\\
T_A & \xrightarrow{\kappa} Y + T_A &T &\xrightarrow{\beta} Y + T \nonumber\\
\{D_A, D_R\} &\xrightarrow{\Bar{\delta}} D & \{T_A, T_R\} &\xrightarrow{\Bar{\delta}} T \nonumber\\
Y& \xrightarrow{\gamma} \emptyset. \label{reactions}
\end{align}
Here, $\gamma$ and $\bar{\delta}$ are the corresponding decay rates constant for the protein and complexes. We assume that transcription factors degrade and dilute irrespective of being bound to a gene. The corresponding reaction rate equations (RREs) can be obtained using mass action kinetics as~\cite{del2014biomolecular}:
\begin{align}
 \Dot{A} & = u_A - d_a^{+} D A + d_a^{-}D_A - t_a^{+} T A + t_a^{-} T_A -\bar{\delta} A \label{sysstr}\\
 \Dot{R} & = u_R - d_r^{+} D R + d_r^{-}D_R - t_r^{+} T R + t_r^{-} T_R -\bar{\delta} R\\
 \Dot{D_A} & = d_a^{+} D A - d_a^{-}D_A - \bar{\delta} D_A\\
 \Dot{D_R} & = d_r^{+} D R - d_r^{-}D_R - \bar{\delta}D_R \\ 
 \Dot{T_A} & = t_a^{+} T A - t_a^{-} T_A - \bar{\delta} T_A \\
 \Dot{T_R} & = t_r^{+} T R - t_r^{-}T_R - \bar{\delta} T_R \\
 \Dot{Y} & =\beta T + \kappa T_A - \gamma Y, \label{eqn:Ydot}
\end{align}
We define the total concentrations of the activator, repressor, target, and decoy species as:
\begin{align}
    A_{\tot} &= A + T_A + D_A, & R_{\tot} = R + T_R + D_R \label{consstr}\\
    T_{\tot} &= T + T_A + T_R, & D_{\tot} = D + D_A + D_R \label{sysend}
\end{align}
with their time derivatives being:
\begin{align}
    \Dot{A_{\tot}} &= u_A - \bar{\delta} A, & \Dot{R_{\tot}} = u_R - \bar{\delta} R \label{at}\\
    \Dot{T_{\tot}} &= 0, & \Dot{D_{\tot}} = 0.
\end{align}

\section{Paradoxical effects at steady state and transients}
For the proposed synthetic circuit governed by equations~(\ref{sysstr})-(\ref{sysend}), the paradoxical effect is captured in two scenarios. First, by varying the levels of activator in the circuit (by changing $u_A$) and second by varying the levels of decoy sites in the system. The knockout of the activator binding to the target is achieved by maintaining a high dissociation constant $k_{ta} = \tfrac{t_a^-}{t_a^+}$. Note: hereon the steady-state concentration of the species $x$ is denoted as $\overline{x}$.

\subsection{Increasing activator causes unintended repression}
\begin{theorem}
\label{tm1}
    The output $\overline{Y}$ of our synthetic circuit depends in a complicated manner on the input $u_A$:
\begin{align}
    & \frac{\mathrm{d} \overline{Y}}{\mathrm{d} u_A} = (\kappa - \beta) \frac{C_1(u_A)}{k_{ta}} + C_2(u_A), \label{t1}
    \end{align}
    where
    \begin{align}
    C_1 &=  C\left(1 + \frac{\kappa \overline{R}}{(\kappa-\beta)k_{tr}} + \frac{\tfrac{T_{\tot}}{k_{tr}}}{1 + \tfrac{\overline{R}}{k_{tr}}+ \tfrac{\overline{A}}{k_{ta}}} +  \frac{\tfrac{D_{\tot}}{k_{dr}} (1 + \tfrac{\overline{A}}{k_{ta}})}{(1 + \tfrac{\overline{R}}{k_{dr}}+ \tfrac{\overline{A}}{k_{da}})^2}\right)\nonumber\\
    C_2 &= \left[ \frac{\kappa}{k_{ta}} - \frac{\beta}{k_{da}}\right] \frac{\overline{R} D_{\tot} C}{k_{tr}k_{dr}(1 + \tfrac{\overline{R}}{k_{dr}}+ \tfrac{\overline{A}}{k_{da}})^2}\nonumber
    \end{align}
with $\overline{A}$ and $\overline{R}$ depend on $u_A$. The full expression for $C$ is provided in Appendix \ref{appA}.
\end{theorem}

\begin{remark}
Notice that $C_1 > 0$ and
\begin{align*}
C_2 \begin{cases}
        > 0 & \text{if }  k_{ta} < \frac{\kappa k_{da}}{\beta}\\
        < 0 & \text{if }  k_{ta} > \frac{\kappa k_{da}}{\beta}.
    \end{cases}
\end{align*}
This means that $ \frac{\mathrm{d} \overline{Y}}{\mathrm{d} u_A}$ is the sum of a positive term and a term which is negative for appropriate parameter choices.
In particular, for large values of $k_{ta}$ the negative second term dominates, so that the dependence of $\overline{Y}$ on $u_A$ is negative:
\begin{align*}
     \frac{\mathrm{d} \overline{Y}}{\mathrm{d} u_A} < 0 \quad \text{for} \quad k_{ta} \rightarrow \infty, 
\end{align*}
contradicting the expectation that $A$ is an activator and hence larger values of $u_A$ should result in larger values of $\overline{Y}$.
\end{remark}


\smallskip

\begin{myproof}
    Using equation (\ref{eqn:Ydot}), the steady-state levels of the output protein is:
    \begin{align}
        \overline{Y} = \frac{\beta \overline{T} + \kappa \overline{T_A}}{\gamma}. \label{ybar}
    \end{align}
At steady state, we have:
    \begin{align}
 \overline{D}_A &= \frac{\overline{D} \overline{A}}{K_{da}}, \qquad \overline{T}_A = \frac{\overline{T} \overline{A}}{K_{ta}},\label{tass}\\
\overline{D}_R &= \frac{\overline{D} \overline{R}}{K_{dr}}, \qquad \overline{T}_R = \frac{\overline{T} \overline{R}}{K_{tr}},
\end{align}
where $K_{xy} = \tfrac{x_y^- + \Bar{\delta}}{x_y^+}$. Note that as $k_{ta} \rightarrow \infty \implies K_{ta} \rightarrow \infty$ for a finite $\Bar{\delta}$. Substituting in equations (\ref{consstr})-(\ref{sysend}), we get:
\begin{align}
    \overline{A_{\tot}} &= \overline{A} + \frac{\overline{T} \overline{A}}{K_{ta}} + \frac{\overline{D} \overline{A}}{K_{da}}, & \overline{R_{\tot}}  = & \overline{R} + \frac{\overline{T} \overline{R}}{K_{tr}} + \frac{\overline{D} \overline{R}}{K_{dr}}, \label{ssstart}\\
    \overline{T_{\tot}} &= \overline{T} + \frac{\overline{T} \overline{A}}{K_{ta}} + \frac{\overline{T} \overline{R}}{K_{tr}}, & \overline{D_{\tot}} =& \overline{D} + \frac{\overline{D} \overline{A}}{K_{da}} + \frac{\overline{D} \overline{R}}{K_{dr}}, \\
    \therefore \overline{T} &= \frac{\overline{T_{\tot}}}{O_t}, & \overline{D} =& \frac{\overline{D_{\tot}}}{O_d}.\label{ssend} \end{align}
 where 
 \begin{align}
    O_t &= 1 + \frac{\overline{A}}{K_{ta}} + \frac{\overline{R}}{K_{tr}}, & O_d =& 1 + \frac{\overline{A}}{K_{da}} + \frac{\overline{R}}{K_{dr}} \nonumber
\end{align}

The paradoxical effect is shown by calculating:
\[\frac{\mathrm{d} \overline{Y}}{\mathrm{d} u_A} = \frac{\beta}{\gamma} \frac{\mathrm{d} \overline{T}}{\mathrm{d} u_A} + \frac{\kappa}{\gamma} \frac{\mathrm{d} \overline{T}_A}{\mathrm{d} u_A} = \frac{\beta}{\gamma \Bar{\delta}} \frac{\mathrm{d} \overline{T}}{\mathrm{d} A_{\tot}} + \frac{\kappa}{\gamma  \Bar{\delta}} \frac{\mathrm{d} \overline{T}_A}{\mathrm{d} A_{\tot}},\]
as $A_{\tot} = \tfrac{u_A}{\Bar{\delta}}$ from equation (\ref{at}). Applying product rule after substituting equation (\ref{tass}) in (\ref{ssend}): 
\[\frac{\mathrm{d} \overline{Y}}{\mathrm{d} u_A} = \frac{1}{\gamma \Bar{\delta}} \left[ \frac{\kappa \overline{A} + \beta K_{ta}}{K_{ta}} \frac{\mathrm{d \overline{T}}}{\mathrm{d} A_{\tot}} + \frac{\kappa \overline{T}}{K_{ta}} \frac{\mathrm{d \overline{A}}}{\mathrm{d} A_{\tot}}\right]\]
Differentiating equation (\ref{ssend}) with $A_{\tot}$ and substituting:
\begin{align}
\frac{\mathrm{d} \overline{Y}}{\mathrm{d} u_A} = \frac{1}{\gamma \Bar{\delta}}& \left (\frac{\kappa \overline{T}}{K_{ta}} - \frac{\kappa \overline{A} + \beta K_{ta}}{K_{ta}} \frac{\overline{T}}{K_{ta} + \overline{A} + \frac{\overline{R} K_{ta}}{K_{tr}}}\right) \frac{\mathrm{d \overline{A}}}{\mathrm{d} A_{\tot}} - \frac{\kappa \overline{A} + \beta K_{ta}}{K_{ta}} \frac{\overline{T}}{K_{tr} + \overline{R} + \frac{\overline{A} K_{tr}}{\gamma \Bar{\delta}K_{ta}}} \frac{\mathrm{d \overline{R}}}{\mathrm{d} A_{\tot}} \label{yinranda}
\end{align}
Calculating each derivative individually using equation (\ref{ssstart}) and substituting in equation (\ref{yinranda}), we get:
\begin{align*}
    \frac{\mathrm{d} \overline{Y}}{\mathrm{d} u_A} = (\kappa - \beta) \frac{C_1}{k_{ta}} + C_2.
\end{align*}
Details of these computations are provided in Appendix \ref{appA}.
\end{myproof}

\begin{figure}[ht]
\centering
\picc{0.7}{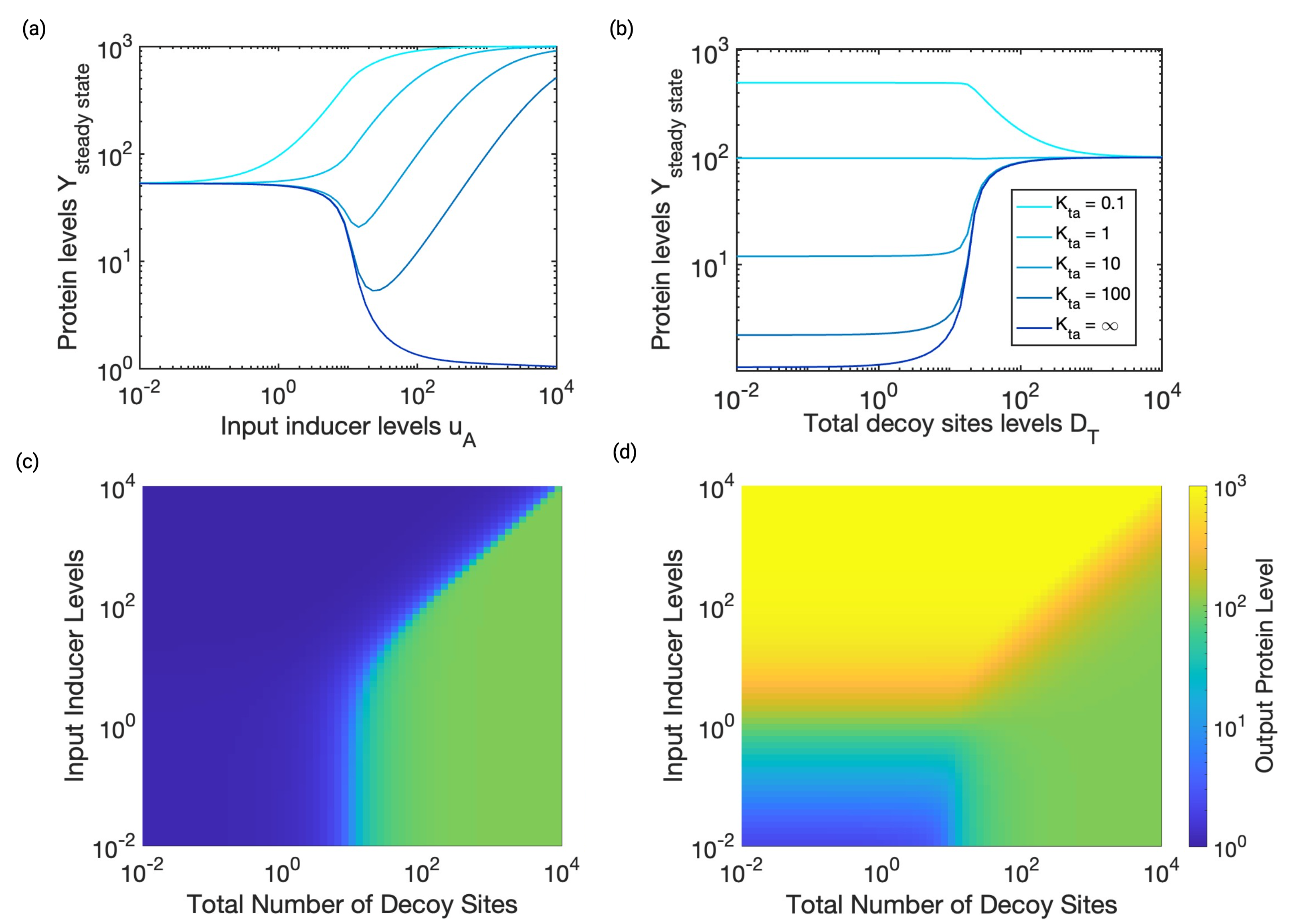}
\caption{Paradoxical effect in steady-state levels of the output protein portrayed for varying the levels of (a) activator input levels ($u_A$ for $D_\tot = 20$) and (b) total decoy sites ($D_{\tot}$ for $u_A = 10$) in the system. Two dimensional response plot between $u_A$ and $D_\tot$ showing the existence of the paradoxical effects for (c) $K_{ta} \rightarrow \infty$ and its absence for (d) $K_{ta} = 0.1$. The parameter values used for the simulation are: $K_{tr} = 0.1, K_{dr} = 0.1, K_{da} = 0.1, T_\tot = 1, R_\tot = 10, \kappa = 1000, \beta = 100,\text{ and } \gamma = 1$.}
\label{fig:ssplots}
\end{figure}

Figure~\ref{fig:ssplots}-a shows the exhibition of paradoxical effect with increasing the input inducer levels of the activator. As $u_A$ is increased, the total amount of activators in the system increases. For low values of $K_{ta}$ ($K_{ta} = 0.1$ and 1), the output protein level increases as expected with an increase in the levels of activator. Increasing $K_{ta}$ (to 10 and 100) and thereby gradually knocking out the direct influence of the activator on the target engenders the paradoxical effect, where increasing $u_A$ leads to an initial decrease in the output levels of $\overline{Y}$ followed by an increase. For the extreme case of complete knockout of the activator with $K_{ta} \rightarrow \infty$, we observe a monotonic decrease in the concentration of the protein. In the two-parameter bifurcation plots in Figure~\ref{fig:ssplots}-c, we see that for high $K_{ta}$ value, the paradoxical affect is observed only after a threshold value of $D_\tot$. For low values of $D_\tot$ (say $10^0$), increasing the activator levels have no significant effects on the output protein levels. On the other hand for high $D_\tot$ (say $10^2$), increasing the activator levels shows a decrease in the output protein levels hence the paradoxical effect. In the case of low $K_{ta}$ value in Figure~\ref{fig:ssplots}-d, we see that increasing $u_A$, increases the protein levels irrespective of the value of $D_{\tot}$.

\begin{corollary}
    The presence of basal expression is necessary for the exhibition of the paradoxical effect.
\end{corollary}
\begin{myproof}
    The presence of basal expression is captured by $\beta$, transcription rate in the inactive/neutral form of the target. Substituting $\beta =0$ in equation (\ref{t1}) implies $C_2 > 0 \forall\ k_{ta}$. Therefore, increasing the amount of activator increases the output levels irrespective of the value of $k_{ta}$.
\end{myproof}

\subsection{Dynamics flip during initial transient \label{Aflip}}
One might wonder at what time can one observe the paradoxical effect. To examine the onset of the paradoxical effect, we determine the small-time relationship between the output $Y$ and the input $u_A$. We can represent the system under consideration as $\dot{x} = f(x,p)$ where $p$ is the set of parameters (i.e., rate constants) that the system depends on and $x = \{ T_R,T_A,D_R,D_A, Y, A_{\tot}, R_{\tot}\}$. We define $Y(x(t))$ as the projection onto the $Y$ species. We indicate the Lie derivative by $L_{\subfxp} Y(x(t)) = (\nabla Y) f(x,p) = \frac{dY}{dt}$.

\begin{theorem}
\label{thm: partials in u_A small time}
    For the proposed synthetic circuit governed by equations (\ref{sysstr})-(\ref{sysend}) with $Y(t,u_A)$ as a function of $t$ and $u_A$, we have:
    \[\partial_{u_A} Y(t, u_A) > 0 \]
    for small enough $t$,
    \[\partial_{u_A} Y(0,u_a) = \partial_{u_a} \Dot{Y}(0,u_a) = \partial_{u_a} \Ddot{Y}(0,u_a) = 0\] and 
    \[ \partial_{u_a} \dddot{Y}(0,u_a) = (\kappa - \beta) t_a^{+}(T_{\tot} - T_R - T_A).\]
\end{theorem}

\begin{myproof}
From equation (\ref{eqn:Ydot}), we have that
\begin{equation}
\label{eq:first lie derivative}
    L_{\subfxp} Y(x(t)) = \beta (T_{\tot} - T_R - T_A) + \kappa T_A - \gamma Y.
\end{equation}
Therefore:
\begin{multline}
\label{eq:second lie derivative}
     L_{\subfxp}(L_{\subfxp} Y(x(t)))\\
     =  \beta(-(t_r^{+}(T_{\tot} - T_R - T_A)(R_{\tot} - D_R - T_R) - t_r^{-}T_R - \bar{\delta} T_R) - (t_a^{+}(T_{\tot} - T_R - T_A)(A_{\tot} - D_A - T_A) - t_a^{-} T_A - \bar{\delta} T_A)) \\+ \kappa (t_a^{+}(T_{\tot} - T_R - T_A)(A_{\tot} - D_A - T_A) - t_a^{-} T_A - \bar{\delta} T_A)  - \gamma (\beta (T_{\tot} - T_R - T_A) + \kappa T_A - \gamma Y)
\end{multline}

Note that upon taking a third lie derivative, only the term of the form $(\kappa - \beta)(t_a^{+}(T_{\tot} - T_R - T_A)(A_{t}))$ will differentiate to produce a term with $u_A$ in it. In particular, we have that
\begin{align*}
    &L_{\subfxp}((\kappa - \beta) t_a^{+}(T_{\tot} - T_R - T_A)(A_{\tot})) \frac{1}{(\kappa - \beta) t_a^{+}} \\
    &=A_{t} L_{\subfxp}(T_{\tot} - T_R - T_A) + (T_{\tot} \!-\! T_R\! - \!T_A)L_{\subfxp}(A_{\tot}) \\
    &=A_{\tot} L_{\subfxp}(T_{\tot} \!- \!T_R\! -\! T_A) + (T_{\tot}\! - T_R\! -\! T_A)(u_A - \bar{\delta} A_{\tot}).
\end{align*}

We see that the expression has a term of the form $(T_{\tot} - T_R - T_A)u_A$. Thus for $ (\kappa - \beta) t_a^{+}(T_{\tot} - T_R - T_A) > 0$, we see that the third lie derivative is nondecreasing in $u_A$. This implies that for small times, $Y$ will be nondecreasing in $u_A$. Indeed, we can write:
\begin{equation}
\label{eq:lie derivative taylor}
Y(t,u_A) = Y(0) + \frac{\dot{Y}(0)}{1!}t^1 + \frac{\ddot{Y}(0)}{2!}t^2 + \frac{\dddot{Y}(0)}{3!}t^3 + O(t^4)
\end{equation}
\[ \implies \partial_{u_A} Y(t, u_A) = \frac{\partial_{u_A}(\dddot{Y}(0))}{3!}t^3 + O(t^4) > 0\]
for small enough $t$.
\end{myproof}

\begin{figure}[ht]
\picc{0.75}{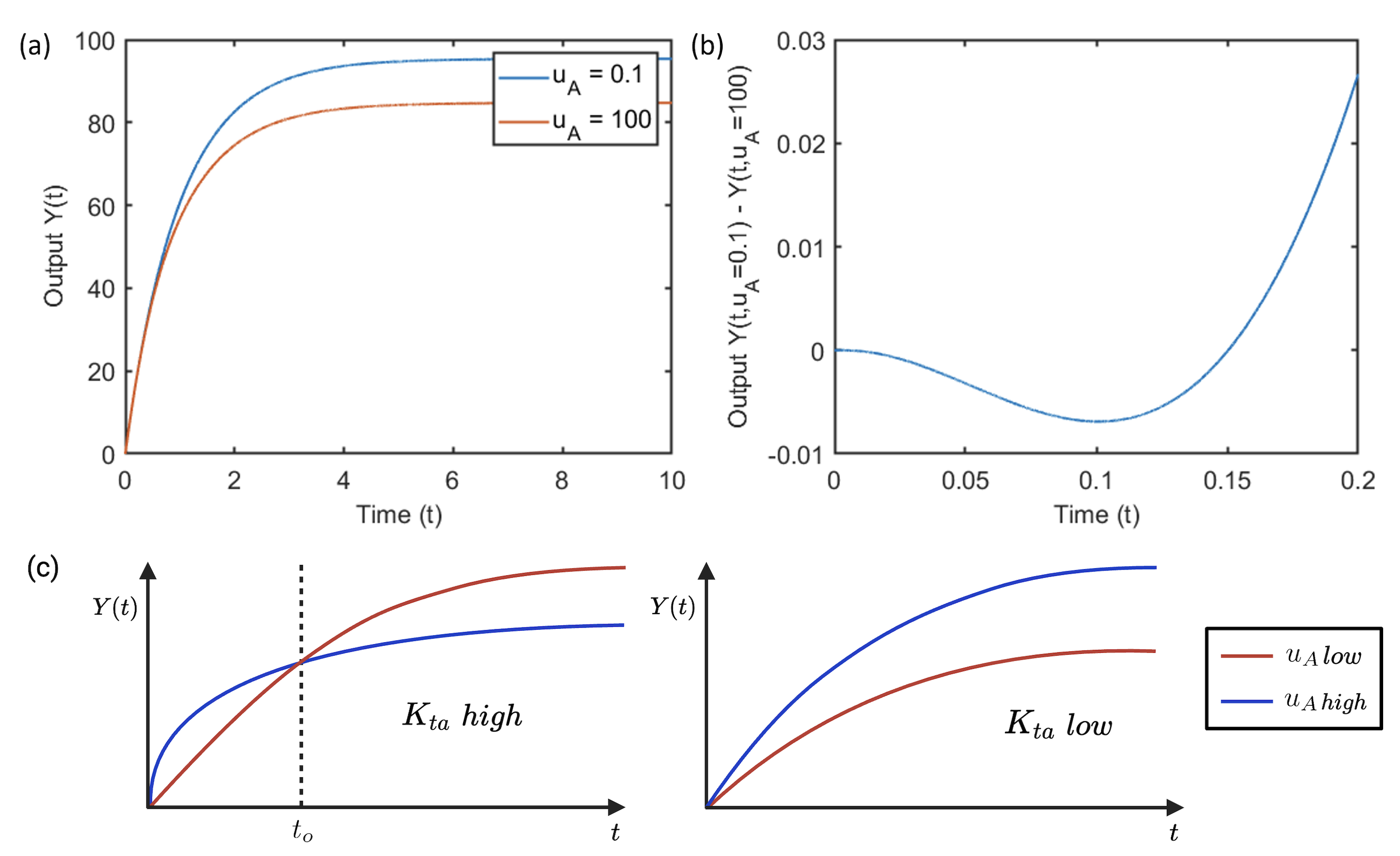}
\caption{Transient dynamics is devoid of paradoxical effects.  Time series of the (a) concentration of the output protein $Y(t)$ for $u_A = 0.1$ and $u_A = 100$, and (b) difference between the output levels of the protein $Y(t,u_A = 0.1) - Y(t, u_A = 100)$, for $K_{ta}=50$. Generally, one may describe the existence for initial crossing times as depicted in (c). The parameter values are same as Figure \ref{fig:ssplots}.} 
\label{fig:transientplots}
\end{figure}
Figure~\ref{fig:transientplots} displays the flip in the behavior of the synthetic circuit after an initial transient to exhibit the paradoxical effect. The time series of $Y(t)$ shows the existence of the paradoxical effect, where the output level for $u_A = 0.1$ is higher than that of $u_A = 100$. However, taking a closer look at the data, the effect emerges after a finite duration of time. This is shown by plotting the difference between the output levels for $u_A = 0.1$ and $u_A = 100$ in Figure~\ref{fig:transientplots}-b. We see that $t<t_o$ (where $t_o>0$ is a threshold value of time for different combinations of inputs, $t_0 \approx 0.15$ for $u_A = 0.1$ and $u_A = 100$), $Y(u_A = 100) > Y(u_A = 0.1)$, i.e., an increase in the activator increases the output protein levels and therefore no paradoxical effect. Whereas for $t>t_o$, $Y(u_A = 100) < Y(u_A = 0.1)$, i.e., an increase in the activator decreases the output protein levels and therefore exhibits the paradoxical effect. Hence, the output protein levels of the circuit with low input values (say $u_A = 0.1$) start at $t = 0$ with a lower initial growth rate, cross the trajectory of an intermediate input level (say $u_A = 100$), before reaching a higher steady state value than the intermediate one, as depicted in Figure~\ref{fig:transientplots}-c. The parameters used for the numerical simulations are somewhat arbitrary, therefore the magnitude of the effects and the timescales could be expected to be different in practice. 

\subsection{Increasing decoy sites increases the output}
The paradoxical effect is also portrayed by varying the levels of decoy sites in the system. 

\begin{theorem}
    The output $\overline{Y}$ of our synthetic circuit depends on $D_\tot$ as follows:
\begin{align}
    & \frac{\mathrm{d} \overline{Y}}{\mathrm{d} D_\tot} = C_3 + \frac{C_4}{K_{ta}} \label{d1}
    \end{align}
    where,
    \begin{align}
    C_3 &= \frac{\frac{C_5}{K_{tr}\gamma}(\frac{\kappa\overline{A}}{K_{ta}} + \beta)\frac{T_\tot}{O^2_t}}{1 + \frac{D_\tot}{K_{dr}O^2_d}(1 + \frac{\overline{A}}{K_{da}}) + \frac{T_\tot}{K_{tr}O^2_t}(1 + \frac{\overline{A}}{K_{ta}})} \nonumber \\
    C_4 &= \frac{\mathrm{d}\overline{A}}{\mathrm{d}D_\tot}\frac{T_\tot}{\gamma O^2_t} \left[\kappa - \beta + \frac{\kappa\overline{R}}{K_{tr}} \right] \nonumber \\
    C_5 &= \frac{\overline{R}}{K_{dr}O_d} - \left[ \frac{\overline{R}D_\tot}{K_{dr}K_{da}O^2_d} + \frac{\overline{R}D_\tot}{K_{tr}K_{ta}O^2_t} \right]\frac{\mathrm{d}\overline{A}}{\mathrm{d}D_\tot} \nonumber
    \end{align} 
with $\frac{\mathrm{d}\overline{A}}{\mathrm{d}D_\tot} < 0$ defined in Appendix \ref{appB} and $\overline{A}$ and $\overline{R}$ depend on $D_\tot$. 
\end{theorem}

\begin{remark}
Notice that $C_3 > 0$ and $C_4 < 0$. This means that $ \frac{\mathrm{d}\overline{Y}}{\mathrm{d} D_\tot}$ is the sum of a positive term and a negative term. 
In particular, for small values of $K_{ta}$ the negative second term dominates, while for large values of $K_{ta}$ the positive first term dominates, such that:
\begin{align*}
     \frac{\mathrm{d} \overline{Y}}{\mathrm{d} D_\tot} > 0 \quad \text{for} \quad k_{ta} \rightarrow \infty, 
\end{align*}
contradicting the expectation that increasing the competitor sites ($D_\tot$) decreases the output $\overline{Y}$.
\end{remark}
\begin{myproof}
Differentiating equation (\ref{ybar}) with respect to $D_\tot$ we have
\[\frac{\mathrm{d}\overline{Y}}{\mathrm{d}D_\tot} = \frac{1}{\gamma} \left[ \frac{\kappa \overline{A}}{K_{ta}} + \beta \right] \frac{\mathrm{d} \overline{T}}{\mathrm{d} D_\tot} +  \frac{1}{\gamma} \frac{\kappa \overline{T}}{K_{ta}} \frac{\mathrm{d} \overline{A}}{\mathrm{d} D_\tot}\]

Using a similar approach to Theorem \ref{tm1}, we compute $\frac{\mathrm{d}\overline{A}}{\mathrm{d}D_\tot}$ and $\frac{\mathrm{d}\overline{T}}{\mathrm{d}D_\tot}$ to get our output in the form (\ref{d1}).
\end{myproof}

Figure~\ref{fig:ssplots}-b shows the variation of steady-state output levels ($\overline{Y}$) as a function of decoy sites ($D_{\tot}$) for different values of $K_{ta}$. For $K_{ta} = 0.1$, we observe the expected behavior of the output levels decreasing as the amount of decoy sites increases. On the contrary, the behavior flips for higher $K_{ta}$ values showing the exhibition of the paradoxical effect. In the two-parameter bifurcation plots in Figure~\ref{fig:ssplots}-c, we see that for high $K_{ta}$ value, the paradoxical affect is observed for all values of $u_A$. Increasing the amount of $D_\tot$, increases the output concentration. However, the value of $D_\tot$ after which the system moves to an activated state depends on the value of $u_A$. Therefore, the onset of the effect can be controlled using the activator levels. In the case of low $K_{ta}$ value in Figure~\ref{fig:ssplots}-d, we see that increasing $D_{\tot}$, decreases the protein levels irrespective of the value of $u_A$.

Next, we examine the onset of the paradoxical effect in a similar manner as section \ref{Aflip}.
\begin{theorem}
Considering the proposed circuit, with $Y$ is a function of $t$ and $D_{\tot}$, i.e. $Y(t, D_{\tot})$, we have:
\[\partial_{D_{\tot}} Y(0,D_{\tot}) = \partial_{D_{\tot}} \dot{Y}(0,D_{\tot}) = \partial_{D_{\tot}} \ddot{Y}(0,D_{\tot}) = 0.\]Then, \[\partial_{D_{\tot}} {\dddot{Y}}(t, D_{\tot}) < 0 \text{ when } At^+_ad^+_a(\kappa - \beta) > Rt^+_rd^+_r(\beta)\] \text{and} \[\partial_{D_{\tot}} \dddot{Y}(t, D_{\tot}) > 0 \text{ when } At^+_ad^+_a(\kappa - \beta) < Rt^+_rd^+_r(\beta)\] for small enough $t$.
\end{theorem}
\begin{myproof}
Similarly from Theorem \ref{thm: partials in u_A small time}, we can repeatedly take Lie derivatives to find dependence on $D_{\tot}$ in the Taylor expansion from equation (\ref{eq:lie derivative taylor}). It follows from equations (\ref{eq:first lie derivative}) and (\ref{eq:second lie derivative}), that the first and second order terms in equation (\ref{eq:lie derivative taylor}) do not have dependence on $D_{\tot}$. 

Looking at the second lie derivative, note that only differentiating $D_A$ or $D_R$ would give terms involving $D_T$. After differentiating the third Lie derivative and only keeping track of these terms, we will see a term of the form
\begin{multline}
    D_{\tot} (-(\kappa - \beta) t_a^+ (T_{\tot} - T_R - T_A) d_a^+ (A_{\tot} - D_A - T_A) \,+ \beta t_r^+ (T_{\tot} - T_R - T_A) d_r^+ (R_{\tot} - D_R - T_R)) \\
    = D_{\tot} (-(\kappa - \beta) t_a^+ d_a^+ T  A + \beta t_r^+ d_r^+ T  R)
\end{multline}
From this term we can observe that if 
\[At^+_ad^+_a(\kappa - \beta) < Rt^+_rd^+_r(\beta)\]
Then for small time, increases in $D_{\tot}$ lead to increases in $Y$, and if 
\[At^+_ad^+_a(\kappa - \beta) > Rt^+_rd^+_r(\beta)\]
for small times, increases in $D_{\tot}$ lead to decreases in $Y$.
\end{myproof}

\section{Additional system properties}

The following claims give some additional information about the qualitative behavior of the system. In particular we can rigorously prove that the system has bounded trajectories, and that there exists a unique equilibrium.

\begin{claim} \label{bounded}
 The proposed synthetic circuit with the reaction network given by equations (\ref{reactions}) admits bounded trajectories.
\end{claim}

\begin{myproof}
    Note we have conservation laws $D + D_A + D_R = D_{\tot}$ and $T + T_A + T_R = T_{\tot}$, which implies all these quantities are bounded above by $e = \max(D_{\tot},T_{\tot})$. Similarly:
    \begin{align*}
        \dot{A} & = -\delta A + u_a - d_a^+ AD - t_a^+ AT + d_a^- D_A + t_a^- T_A  \leq -\delta A  - d_a^+ AD - t_a^+ AT + (d_a^- + t_a^-) e + u_a.
    \end{align*}   
    Thus for large enough values of $A$, the above expression will be negative, therefore $A$ is bounded from above. The same reasoning applies to $R$:
    \begin{align*}
        \dot{R} = -\delta R + u_R - d_r^+ RD - t_r^+ RT + d_r^- D_R + t_r^- T_R 
        \leq -\delta R  + (d_r^- + t_r^-) e + u_R
    \end{align*}
     In particular, the polytope described by the equations:
    \[D + D_A + D_R = D_{\tot}, \qquad T + T_A + T_R = T_{\tot},\]
    \[A \leq \frac{(d_a^- + t_a^-)e + u_a}{\delta}, \qquad R \leq \frac{(d_r^- + t_r^-) e + u_R}{\delta},\]
    \[A, R, D_A, D_R, T_A, T_R, D, T \geq 0\]
    is invariant under our vector field, and thus our trajectories are bounded.
\end{myproof}
\begin{claim}
    The reaction network has an equilibrium.
\end{claim}
\begin{myproof}
    From Claim \ref{bounded}, there is a compact and convex set of values of our system species $\{T_R,T_A,D_R,D_A, Y, A_{\tot}, R_{\tot}\}$ that is closed under the time evolution of our system. By Brouwer's fixed point theorem for every $t \geq 0$ we have the time evolution $\phi_t$ has a fixed point $\overline{x}_t$. Take a sequence of $t_n \rightarrow 0$, the sequence $\overline{x}_{t_n}$ has a limit point $\overline{x}$. If $\overline{x}$ was not an equilibrium, then for small $t$ it would not be a fixed point for $\phi_t$, which is a contradiction. Thus $\overline{x}$ is an equilibrium of our vector field. 
\end{myproof}

This theorem applies to our system, and implies that our system has at least one equilibrium. Next we will note that this equilibrium must be in fact unique.



In order to know that an equilibrium is in fact the unique equlibrium of the system, we can use the notion of injectivity. We say the system is \textit{injective}, as in \cite{doi:10.1137/120873388}, if for all possible kinetic parameters the system $\dot{x} = f(x,p)$ is such that $f(x,p)$ is an injective function, no matter the choice of $p$. If this is true, it implies $f(x,p) = 0$ has at most one solution (i.e., we have at most one equilibrium).

\begin{claim}
    The proposed synthetic circuit is injective.
\end{claim}
\begin{myproof}
    Using the conservation laws, $T_{\tot} = T + T_A + T_R$ and $D_{\tot} = D + D_A + D_R$, we can form the ``extended rate function" by simply replacing the dynamics for $\dot{T}$ and $\dot{D}$ with our conservation laws. We can then take the determinant of the Jacobian of this modified rate function and verify all its terms are positive, which implies injectivity by Theorem 8.1 in \cite{doi:10.1137/120873388}. Mathematica code for this computation can be found in~\cite{al2024competition}. 
Inspection of the output confirms the Jacobian is always nonzero and thus our system is injective.
\end{myproof}

\section{Discussion}

Starting from a hypothesis concerning the role of off-target binding in explaining paradoxical behaviors of activators and repressors observed experimentally in cancer cell culture experiments, this paper proposed a synthetic biology experiment to reproduce these behaviors in a controlled situation. A mathematical analysis was used to identify appropriate parameter regimes, and theoretical results were obtained. Work is ongoing to build the appropriate synthetic constructs and perform confirmatory experiments. Model-driven experiments are being performed using CRISPRa as the activator and CRISPRi as the repressor.

\subsection*{Acknowledgements}
The authors wish to thank Dr. Polly Yu for very useful discussions and suggestions regarding the use of the injectivity property.

 \bibliographystyle{IEEEtran}
\bibliography{2024_cdc_competition_repressor_activator.bib}{}

\newpage
    \appendix
    
\subsection{Calculating the derivatives $\frac{d\overline{R}}{d\overline{A}_{\tot}}, \frac{d\overline{A}}{d\overline{A}_{\tot}}$ and thereby $\frac{dY}{d\overline{A}_{\tot}}$}
\label{appA}
Using equation (\ref{ssstart}) - (\ref{ssend}),  
\[\overline{R_{\tot}} - \overline{R} + \frac{\overline{T} \overline{R}}{K_{tr}} + \frac{\overline{D} \overline{R}}{K_{dr}} = 0 \]
Taking the derivative with respect to $\overline{A}_{\tot}$:
\begin{align*}
    \frac{d\overline{R}}{d\overline{A}_{\tot}}[
1 + \frac{D_{\tot}}{K_{dr}(1 + \frac{\overline{R}}{K_{dr}} + \frac{\overline{A}}{K_{da}})} + &\frac{T_{\tot}}{K_{tr}(1 + \frac{\overline{R}}{K_{tr}} + \frac{\overline{A}}{K_{ta}})} ] - \frac{\overline{R}D_{\tot}}{K_{dr}(O_d)^2}(\frac{1}{K_{dr}}\frac{d\overline{R}}{d\overline{A}_{\tot}} + \frac{1}{K_{da}}\frac{d\overline{A}}{d\overline{A}_{\tot}}) \\
- \frac{\overline{R}T_{\tot}}{K_{tr}(O_t)^2}(\frac{1}{K_{tr}}\frac{d\overline{R}}{d\overline{A}_{\tot}} + \frac{1}{K_{ta}}\frac{d\overline{A}}{d\overline{A}_{\tot}}) = 0 
\end{align*}

After reorganizing, we get:
\[ \frac{d\overline{R}}{d\overline{A}_{\tot}} = \frac{(\frac{\overline{R}D_{\tot}}{K_{da}K_{dr}(1 + \frac{\overline{R}}{K_{dr}} + \frac{\overline{A}}{K_{da}})^2} + \frac{\overline{R}T_{\tot}}{K_{ta}K_{tr}(1 + \frac{\overline{R}}{K_{tr}} + \frac{\overline{A}}{K_{ta}})^2}) }
{(1 + \frac{D_{\tot}}{K_{dr}(1 + \frac{\overline{R}}{K_{dr}} + \frac{\overline{A}}{K_{da}})} (1 - \frac{\frac{\overline{R}}{K_{dr}}}{(1 + \frac{\overline{R}}{K_{dr}} + \frac{\overline{A}}{K_{da}})}) \frac{T_{\tot}}{K_{tr}(1 + \frac{\overline{R}}{K_{tr}} + \frac{\overline{A}}{K_{ta}})}(1 - \frac{\frac{\overline{R}}{K_{tr}}}{(1 + \frac{\overline{R}}{K_{tr}} + \frac{\overline{A}}{K_{ta}})}) )} \frac{d\overline{A}}{d\overline{A}_{\tot}}\]

Doing a similar approach on the conservation law for $\overline{A}_\tot$, we get:
\begin{align*}
\frac{d\overline{A}}{d\overline{A}_{\tot}}&(1 + \frac{D_{\tot}}{K_{da}(1 + \frac{\overline{R}}{K_{dr}} + \frac{\overline{A}}{K_{da}})} + \frac{T_{\tot}}{K_{ta}(1 + \frac{\overline{R}}{K_{tr}} + \frac{\overline{A}}{K_{ta}})}) \\
&- \frac{\overline{A}D_{\tot}}{K_{da}(O_t)^2}(\frac{1}{K_{dr}}\frac{d\overline{R}}{d\overline{A}_{\tot}} + \frac{1}{K_{da}}\frac{d\overline{A}}{d\overline{A}_{\tot}}) 
- \frac{\overline{A}T_{\tot}}{K_{ta}(O_t)^2}(\frac{1}{K_{tr}}\frac{d\overline{R}}{d\overline{A}_{\tot}} + \frac{1}{K_{ta}}\frac{d\overline{A}}{d\overline{A}_{\tot}}) = 1
\end{align*}

Rearranging, we get:

\begin{align*}
    \frac{d\overline{A}}{d\overline{A}_{\tot}}(1 + \frac{D_{\tot}}{K_{da}(1 + \frac{\overline{R}}{K_{dr}} + \frac{\overline{A}}{K_{da}})}  &(1 - \frac{\frac{\overline{A}}{K_{da}}}{1 + \frac{\overline{R}}{K_{dr}} + \frac{\overline{A}}{K_{da}}}) + \frac{T_{\tot}}{K_{ta}(1 + \frac{\overline{R}}{K_{tr}} + \frac{\overline{A}}{K_{ta}})} (1 - \frac{\frac{\overline{A}}{K_{ta}}}{1 + \frac{\overline{R}}{K_{tr}} + \frac{\overline{A}}{K_{ta}}}))\\
    & = 1 + (\frac{\overline{A}D_{\tot}}{K_{da}K_{dr}(1 + \frac{\overline{R}}{K_{dr}} + \frac{\overline{A}}{K_{da}})^2} + \frac{\overline{A}T_{\tot}}{K_{ta}K_{tr}(1 + \frac{\overline{R}}{K_{tr}} + \frac{\overline{A}}{K_{ta}})^2}) \frac{d\overline{R}}{d\overline{A}_{\tot}}
\end{align*}

Substituting $\frac{d\overline{R}}{d\overline{A}_{\tot}}$:
\[\frac{d\overline{A}}{d\overline{A}_{\tot}} = \frac{(1 + \frac{D_{\tot}(1 + \frac{\overline{A}}{K_{da}})}{K_{dr}(1 + \frac{\overline{R}}{K_{dr}} + \frac{\overline{A}}{K_{da}})^2} + \frac{T_{\tot}(1 + {\frac{\overline{A}}{K_{ta}}})}{K_{tr}(1 + \frac{\overline{R}}{K_{tr}} + \frac{\overline{A}}{K_{ta}})^2})}{\text{De}}\]

Where:
\begin{align*}
    \text{De} &= 1 + \frac{D_{\tot}(1 + \frac{\overline{A}}{K_{da}})}{K_{dr}(1 + \frac{\overline{R}}{K_{dr}} + \frac{\overline{A}}{K_{da}})^2} + \frac{T_{\tot}(1 + {\frac{\overline{A}}{K_{ta}}})}{K_{tr}(1 + \frac{\overline{R}}{K_{tr}} + \frac{\overline{A}}{K_{ta}})^2} + \frac{D_{\tot}(1 + \frac{\overline{R}}{K_{dr}})}{K_{da}(1 + \frac{\overline{R}}{K_{dr}} + \frac{\overline{A}}{K_{da}})^2} \\
    & \qquad \qquad+ \frac{D_{\tot}^2}{K_{dr}K_{da}(1 + \frac{\overline{R}}{K_{dr}} + \frac{\overline{A}}{K_{da}})^4}(1 + \frac{\overline{R}}{K_{dr}} + \frac{\overline{A}}{K_{da}}) + \frac{D_{\tot}(1 + \frac{\overline{R}}{K_{dr}} + \frac{\overline{A}}{K_{ta}})T_{\tot}}{K_{da}(1 + \frac{\overline{R}}{K_{dr}} + \frac{\overline{A}}{K_{da}})^2K_{tr}(1 + \frac{\overline{R}}{K_{tr}} + \frac{\overline{A}}{K_{ta}})^2} \\
    & \qquad \qquad + \frac{T_{\tot}(1 + \frac{\overline{R}}{K_{tr}})}{K_{ta}(1 + \frac{\overline{R}}{K_{tr}} + \frac{\overline{A}}{K_{ta}})^2} + \frac{T_{\tot}(1 + \frac{\overline{R}}{K_{tr}} + \frac{\overline{A}}{K_{da}})D_{\tot}}{K_{ta}(1 + \frac{\overline{R}}{K_{tr}} + \frac{\overline{A}}{K_{ta}})^2K_{dr}(1 + \frac{\overline{R}}{K_{dr}} + \frac{\overline{A}}{K_{da}})^2} + \frac{T_{\tot}^2(1 + \frac{\overline{R}}{K_{tr}} + \frac{\overline{A}}{K_{ta}})}{K_{tr}K_{ta}(1 + \frac{\overline{R}}{K_{tr}} + \frac{\overline{A}}{K_{ta}})^4}
\end{align*}

Now that we have explicit equations for the derivatives, $\frac{d\overline{A}}{d\overline{A}_{\tot}}$ and $\frac{d\overline{R}}{d\overline{A}_{\tot}}$, we look at $\frac{dY}{d\overline{A}_{\tot}}$:

\begin{align*}
    \frac{dY}{d\overline{A}_{\tot}} & = \frac{1}{\gamma}\left(\frac{\kappa}{K_{ta}}\overline{A} + \beta\right)\frac{dT}{d\overline{A}_{\tot}} +\frac{1}{\gamma}\frac{\kappa T}{K_{ta}} \frac{d\overline{A}}{d\overline{A}_{\tot}}\\
   & = \frac{1}{\gamma}\frac{\kappa T}{K_{ta}}\frac{d\overline{A}}{d\overline{A}_{\tot}} - \frac{1}{\gamma}(\frac{\kappa}{K_{ta}}\overline{A} + \beta)\frac{T_{\tot}}{(1 + \frac{\overline{R}}{K_{tr}} + \frac{\overline{A}}{K_{ta}})^2} \left(\frac{1}{K_{tr}} \frac{d\overline{R}}{d\overline{A}_{\tot}} + \frac{1}{K_{ta}}\frac{d\overline{A}}{d\overline{A}_{\tot}}\right)\\
    &= \frac{1}{\gamma \text{De}}(\frac{\kappa T_{\tot}}{K_{ta}(1 + \frac{\overline{R}}{K_{tr}} + \frac{\overline{A}}{K_{ta}})}(1 + \frac{D_{\tot}(1 + \frac{\overline{A}}{K_{da}})}{K_{dr}(1 + \frac{\overline{R}}{K_{dr}} + \frac{\overline{A}}{K_{da}})^2} + \frac{T_{\tot}(1 + \frac{\overline{A}}{K_{ta}})}{K_{tr}(1 + \frac{\overline{R}}{K_{tr}} + \frac{\overline{A}}{K_{ta}})^2}) \\
    & \qquad \qquad - (\frac{\kappa}{K_{ta}}\overline{A} + \beta)\frac{T_{\tot}}{(1 + \frac{\overline{R}}{K_{tr}} + \frac{\overline{A}}{K_{ta}})^2} (\frac{1}{K_{tr}} (\frac{\overline{R}D_{\tot}}{K_{da}K_{dr}(1 + \frac{\overline{R}}{K_{dr}} + \frac{\overline{A}}{K_{da}})^2} + \frac{\overline{R}T_{\tot}}{K_{ta}K_{tr}(1 + \frac{\overline{R}}{K_{tr}} + \frac{\overline{A}}{K_{ta}})^2}) \\
    & \qquad \qquad + \frac{1}{K_{ta}} (1 + \frac{D_{\tot}(1 + \frac{\overline{A}}{K_{da}})}{K_{dr}(1 + \frac{\overline{R}}{K_{dr}} + \frac{\overline{A}}{K_{da}})^2})))\\
    &= \frac{1}{\gamma\text{De}}(\frac{\kappa T_{\tot}}{K_{ta}(1 + \frac{\overline{R}}{K_{ta}} + \frac{\overline{A}}{K_{ta}})^2}(1 + \frac{D_{\tot}(1 + \frac{\overline{A}}{K_{da}})}{K_{dr}(1 + \frac{\overline{R}}{K_{dr}} + \frac{\overline{A}}{K_{da}})^2} + \frac{T_{\tot}(1 + \frac{\overline{A}}{K_{ta}})}{K_{tr}(1 + \frac{\overline{R}}{K_{tr}} + \frac{\overline{A}}{K_{ta}})^2}) (1 + \frac{\overline{R}}{K_{tr}}) \\
   & \qquad \qquad - \frac{\overline{A}}{K_{tr}}(\frac{\kappa}{K_{ta}}\overline{A} + \beta) \frac{T_{\tot}}{(1 + \frac{\overline{R}}{K_{tr}} + \frac{\overline{A}}{K_{ta}})^2} \frac{1}{K_{tr}}(\frac{\overline{R}D_{\tot}}{K_{da}K_{dr}(1 + \frac{\overline{R}}{K_{dr}} + \frac{\overline{A}}{K_{da}})^2} + \frac{\overline{R}T_{\tot}}{K_{ta}K_{tr}(1 + \frac{\overline{R}}{K_{tr}} + \frac{\overline{A}}{K_{ta}})^2}) \\
    & \qquad \qquad- \frac{\beta T_{\tot}}{(1 + \frac{\overline{R}}{K_{tr}} + \frac{\overline{A}}{K_{ta}})^2} \frac{1}{K_{ta}}(1 + \frac{D_{\tot}(1 + \frac{\overline{A}}{K_{da}})}{K_{dr}(1 + \frac{\overline{R}}{K_{dr}} + \frac{\overline{A}}{K_{da}})^2} + \frac{T_{\tot}(1 + \frac{\overline{A}}{K_{ta}})}{K_{tr}(1 + \frac{\overline{R}}{K_{tr}} + \frac{\overline{A}}{K_{ta}})^2}))
\end{align*}

Therefore:
\begin{align*}
    &\frac{dY}{d\overline{A}_{\tot}} = \frac{C}{\gamma} \frac{T_{\tot}}{(1 + \frac{\overline{R}}{K_{tr}} + \frac{\overline{A}}{K_{ta}})}(\frac{\kappa}{K_{ta}}(1 + \frac{\overline{R}}{K_{tr}} + \frac{D_{\tot}(1 + \frac{\overline{A}}{K_{da}} + \frac{\overline{R}}{K_{tr}})}{K_{dr}(1 + \frac{\overline{R}}{K_{dr}} + \frac{\overline{A}}{K_{da}})^2}  + \frac{T_{\tot}}{K_{tr}(1 + \frac{\overline{R}}{K_{tr}} + \frac{\overline{A}}{K_{ta}})}) \\
    & \qquad \qquad - \beta(\frac{\overline{R}\frac{D_{\tot}}{K_{tr}}}{K_{da}K_{dr}(1 + \frac{\overline{R}}{K_{dr}} + \frac{\overline{A}}{K_{da}})^2} + \frac{\overline{R}\frac{T_{\tot}}{K_{tr}}}{K_{ta}K_{tr}(1 + \frac{\overline{R}}{K_{tr}} + \frac{\overline{A}}{K_{ta}})^2} + \frac{1}{K_{ta}} \\
    & \qquad\qquad+ \frac{D_{\tot}(1 + \frac{\overline{A}}{K_{da}})}{K_{ta}K_{dr}(1 + \frac{\overline{R}}{K_{dr}} + \frac{\overline{A}}{K_{da}})^2}  \frac{T_{\tot}(1 + \frac{\overline{A}}{K_{ta}})}{K_{ta}K_{tr}(1 + \frac{\overline{R}}{K_{tr}} + \frac{\overline{A}}{K_{ta}})^2} ))
\end{align*}
where $C = \frac{1}{\text{De}}$.

\newpage

\subsection{Calculating the derivatives $\frac{d\overline{Y}}{D_{\tot}}$, $\frac{d\overline{R}}{dD_{\tot}}$, $\frac{d\overline{A}}{dD_{\tot}}$}
\label{appB}
\begin{align*}
    &\frac{d\overline{Y}}{D_{\tot}} = \frac{1}{\gamma}(\kappa\frac{dT_A}{dD_{\tot}} + \beta\frac{dT}{dD_{\tot}})
\end{align*}

Rearranging, we get:
\begin{align*}
    \frac{d\overline{Y}}{d D_{\tot}} & = \frac{1}{\gamma}(\frac{\kappa \overline{A}}{K_{ta}} + \beta)\frac{dT}{dD_{\tot}} + \frac{1}{\gamma}\frac{\kappa T}{K_{ta}}\frac{d\overline{A}}{dD_{\tot}} =\frac{1}{\gamma}(\frac{\kappa \overline{A}}{K_{ta}} + \beta)\frac{dT}{dD_{\tot}} + \frac{1}{\gamma}\frac{\kappa T}{K_{ta}}\frac{d\overline{A}}{dD_{\tot}}
\end{align*}

Now, differentiating with respect to $D_{\tot}$ we have:

\begin{align*}
    &\frac{dT}{dD_{\tot}} = -\frac{T_{\tot}}{O^2_t}(\frac{1}{K_{tr}}\frac{d\overline{R}}{dD_{\tot}} + \frac{1}{K_{ta}}\frac{d\overline{A}}{dD_{\tot}})    
\end{align*}

Differentiating the conservation law for $R$ we get:

\begin{align*}
    &\frac{d\overline{R}}{dD_\tot}(1 + \frac{D_{\tot}}{O_d}K_{dr} + \frac{T_{\tot}}{O_t}K_{tr}) - \frac{\overline{R}D_{\tot}}{K_{dr}O^2_d}(\frac{1}{K_{dr}\frac{d\overline{R}}{dD_{\tot}} + K_{da}\frac{d\overline{A}}{dD_{\tot}}}) + \frac{\overline{R}}{K_{dr}O_d} - \frac{\overline{R}_{T_{\tot}}}{K_{tr}O^2_t}(\frac{1}{K_{tr}}\frac{d\overline{R}}{dD_{\tot}} + \frac{1}{K_{ta}}\frac{d\overline{A}}{dD_{\tot}}) = 0 \\
    &\frac{d\overline{R}}{dD_\tot}(1 + \frac{D_{\tot}}{O_d}K_{dr} + \frac{T_{\tot}}{O_t}K_{tr} - \frac{\overline{R}D_{\tot}}{K^2{dr}O_d^2} - \frac{\overline{R}T_{\tot}}{K^2{tr}O_t^2}) + \frac{\overline{R}}{K_{dr}O_d}  = (\frac{\overline{R}D_{\tot}}{K_{da}K_{dr}O^2_d} + \frac{\overline{R}T_{\tot}}{K_{ta}K_{tr}O^2_t})\frac{d\overline{A}}{d\overline{A}_{\tot}}
\end{align*}

Similarly for $\overline{A}$ we get:

\begin{align*}
    & \frac{d\overline{A}}{dD_{\tot}}(1 + \frac{D_{\tot}}{K_{da}O_d} + \frac{T_{\tot}}{K_{ta}O_t}) - \frac{\overline{A}D_{\tot}}{K_{da}O^2_d}(\frac{1}{K_{dr}}\frac{d\overline{R}}{dD_{\tot}} + \frac{1}{K_{da}}\frac{d\overline{A}}{dD_{\tot}}) + \frac{\overline{A}}{K_{da}O_d)} - \frac{\overline{A}T_{\tot}}{K_{ta}O^2_t}(\frac{1}{K_{tr}}\frac{d\overline{R}}{dD_{\tot}} + \frac{1}{K_{ta}}\frac{d\overline{A}}{dD_{\tot}}) = 0 \\
    & \frac{d\overline{A}}{dD_{\tot}}(1 + \frac{D_{\tot}}{K_{da}(O_d)}(1 - \frac{\overline{A}}{K_{da}O_d}) + \frac{T_{\tot}}{K_{ta}(O_t)(1 - \frac{\overline{A}}{K_{ta}O_t})}) + \frac{\overline{A}}{K_{da}O_d} \\
    & \hspace{30ex} = (\frac{\overline{A}D_{\tot}}{K_{da}K_{dr}O^2_d} + \frac{\overline{A}T_{\tot}}{K_{ta}K_{tr}O^2_t})\frac{d\overline{R}}{dD_{\tot}} \\
    & \hspace{30ex} = (\frac{\overline{A}D_{\tot}}{K_{da}K_{dr}O^2_d} + \frac{\overline{A}T_{\tot}}{K_{ta}K_{tr}O^2_t})\frac{(\frac{\overline{R}D_{\tot}}{K_{da}K_{dr}O^2_d} + \frac{\overline{R}T_{\tot}}{K_{ta}K_{tr}O^2_t})\frac{d\overline{A}}{d\overline{A}_{\tot}} - \frac{\overline{R}}{K_{dr}O_d}}{1 + \frac{D_{\tot}}{K_{dr}(O_d)}(1 - \frac{\overline{R}}{K_{dr}O_d}) + \frac{T_{\tot}}{K_{tr}(O_t)(1 - \frac{\overline{A}}{K_{tr}O_t})}}
\end{align*}

Substituting these in we can get:

\begin{align*}
    \frac{d\overline{A}}{dD_{\tot}} = \frac{(1 + \frac{D_{\tot}(1 + \frac{\overline{A}}{K_{da}})}{K_{dr}O^2_d} + \frac{T_{\tot}(1 + \frac{\overline{A}}{K_{ta}})}{K_{tr}O^2_t})(\frac{-\overline{A}}{K_{da}O_d}) - (\frac{\overline{A}D_{\tot}}{K_{da}K_{dr}O^2_d} + \frac{\overline{A}T_{\tot}}{K_{ta}K_{tr}O^2_t})\frac{\overline{R}}{K_{dr}O_d}}{1 + \frac{D_{\tot}(1 + \frac{\overline{A}}{K_{da}})}{K_{dr}O^2_d} + \frac{T_{\tot}(1 + \frac{\overline{A}}{K_{ta}} + \frac{T_\tot}{K_{ta} O_t})}{K_{tr}O^2_t} + \frac{D_{\tot}(1 + \frac{\overline{R}}{K_{dr}} + \frac{D_\tot}{K_{dr} O_d})}{K_{da}O^2_d} + \frac{D_{\tot}(1 + \frac{\overline{R}}{K_{dr}} + \frac{\overline{A}}{K_{ta}})T_{\tot}}{K_{da}O^2_d K_{tr}O^2_t} + \frac{T_{\tot}(1 + \frac{\overline{R}}{K_{tr}})}{K_{ta}(O^2_t)} + \frac{T_{\tot}(1 + \frac{\overline{R}}{K_{tr}} + \frac{\overline{A}}{K_{da}})D_{\tot}}{K_{ta}O^2_t K_{dr}O^2_d}}
\end{align*}

Once again we have derived explicit equations for $\frac{d\overline{R}}{dD_{\tot}}$, $\frac{d\overline{A}}{dD_{\tot}}$, we compute $\frac{d\overline{Y}}{dD_{\tot}}$.

Notice from the rearranged form, we can factor out $K_{ta}$ into a form such we have one portion with terms from $\frac{d\overline{T}}{dD_{\tot}}$ and $\frac{d\overline{A}}{dD_{\tot}}$ in the numerator and $K_{ta}$ in the denominator, and the other portion a function of variables independent of $K_{ta}$. Substituting and rearranging our closed form derivatives into $\frac{d\overline{Y}}{dD_{\tot}}$, we get:

\begin{align*}
    \frac{\mathrm{d} \overline{Y}}{\mathrm{d} D_\tot} = \frac{1}{K_{tr}\gamma}\frac{\frac{\overline{R}}{K_{dr}O_d} - \left[ \frac{\overline{R}D_\tot}{K_{dr}K_{da}O^2_d} + \frac{\overline{R}D_\tot}{K_{tr}K_{ta}O^2_t} \right]\frac{\mathrm{d}\overline{A}}{\mathrm{d}D_\tot}(\frac{\kappa\overline{A}}{K_{ta}} + \beta)\frac{T_\tot}{O^2_t}}{1 + \frac{D_\tot}{K_{dr}O^2_d}(1 + \frac{\overline{A}}{K_{da}}) + \frac{T_\tot}{K_{tr}O^2_t}(1 + \frac{\overline{A}}{K_{ta}})} + \frac{\frac{\mathrm{d}\overline{A}}{\mathrm{d}D_\tot}\frac{T_\tot}{\gamma O^2_t} \left[\kappa - \beta + \frac{\kappa\overline{R}}{K_{tr}} \right]}{K_{ta}}
\end{align*}

\subsection{Mathematica Code}
\label{appC}
\begin{lstlisting}[language=Mathematica,caption={Code for injectivity proof}]
Adot[A_, R_, Da_, Dr_, Ta_, Tr_, t_, d_] = -k1 A*D + k2 Da - k5 A*T + k6 Ta - k9  A; 

Rdot[A_, R_, Da_, Dr_, Ta_, Tr_, T_, D_] = -k3 R*D + k4 Dr - k7 R*T +    k8 Tr - k10  R; 
 
Dadot[A_, R_, Da_, Dr_, Ta_, Tr_, T_, D_] = k1 A*D - k2 Da - k13 Da  ; 

Drdot[A_, R_, Da_, Dr_, Ta_, Tr_, T_, D_] =   k3 R*D - k4 Dr - k14 Dr  ;

Tadot[A_, R_, Da_, Dr_, Ta_, Tr_, T_, D_] = k5 A*T - k6 Ta - k11 Ta ;

Trdot[A_, R_, Da_, Dr_, Ta_, Tr_, T_, D_] =   k7 R*T - k8 Tr - k12 Tr  ;

Tdot[A_, R_, Da_, Dr_, Ta_, Tr_, T_, D_] = -k5 A*T + k6 Ta - k7 R*T +    k8 Tr + k11 Ta + k12 Tr  ;

Ddot[A_, R_, Da_, Dr_, Ta_, Tr_, T_, D_] = -k1 A*D + k2 Da - k3 R*D +    k4 Dr + k13 Da + k14 Dr ;

ConsvT[A_, R_, Da_, Dr_, Ta_, Tr_, T_, D_] = Tr + Ta + T; 

ConsvD[A_, R_, Da_, Dr_, Ta_, Tr_, T_, D_] = Dr + Da + D ; 

modRHS = {Adot[A, R, Da, Dr, Ta, Tr, T, D],
   Rdot[A, R, Da, Dr, Ta, Tr, T, D],
   Dadot[A, R, Da, Dr, Ta, Tr, T, D],
   Drdot[A, R, Da, Dr, Ta, Tr, T, D],
   Tadot[A, R, Da, Dr, Ta, Tr, T, D],
   Trdot[A, R, Da, Dr, Ta, Tr, T, D],
   ConsvT[A, R, Da, Dr, Ta, Tr, T, D],
   ConsvD[A, R, Da, Dr, Ta, Tr, T, D]
    }; 
    
Grad[modRHS, {A, R, Da, Dr, Ta, Tr, T, D}] // Det

\end{lstlisting}

\end{document}